\documentclass[12pt]{iopart}
\usepackage{url}

\usepackage{graphicx} % For inserting images
\usepackage{tabularx}
\usepackage{iopams}
\graphicspath{ {./images/} }
\usepackage{caption}
\usepackage{float} % To place figures and tables exactly
\usepackage{xcolor}
\usepackage{natbib}
\bibpunct[, ]{(}{)}{,}{a}{}{,}%
\usepackage{multirow}
\usepackage{lscape}

\begin{document}

\title{New Insights into Automatic Treatment Planning for Cancer Radiotherapy Using Explainable Artificial Intelligence}

\author{Md Mainul Abrar\textsuperscript{1}, Xun Jia\textsuperscript{2}, Yujie Chi\textsuperscript{1}}

\address{\textsuperscript{1}Department of Physics, The University of Texas at Arlington, Arlington, TX, United States.\\
\textsuperscript{2}Department of Radiation Oncology and Molecular Radiation Sciences, Johns Hopkins University, Baltimore, MD, United States.}
\ead{Yujie.Chi@uta.edu}
\vspace{10pt}

\begin{abstract}

Objective: This study aims to uncover the opaque decision-making process of an artificial intelligence (AI) agent for automatic treatment planning.

Approach: We examined a previously developed AI agent based on the Actor-Critic with Experience Replay (ACER) network, which automatically tunes treatment planning parameters (TPPs) for inverse planning in prostate cancer intensity modulated radiotherapy. We selected multiple checkpoint ACER agents from different stages of training and applied an explainable AI (EXAI) method to analyze the attribution from dose–volume histogram (DVH) inputs to TPP-tuning decisions. We then assessed each agent’s planning efficacy and efficiency, and evaluated their policy space and final TPP tuning space. Combining findings from these approaches, we systematically examined how ACER agents generated high-quality treatment plans in response to different DVH inputs.

Main Results: Attribution analysis revealed that ACER agents progressively learned to identify dose-violation regions from DVH inputs and promote appropriate TPP-tuning actions to mitigate them. Organ-wise similarities between DVH attributions and dose-violation reductions ranged from 0.25 to 0.5 across tested agents. While all agents achieved comparably high final planning scores, their planning efficiency and stability differed. Agents with stronger attribution–violation similarity required fewer tuning steps (~12–13 vs. 22), exhibited a more concentrated TPP-tuning space with lower entropy (~0.3 vs. 0.6), converged on adjusting only a few key TPPs, and showed smaller discrepancies between practical tuning steps and the theoretical steps needed to move from initial values to the final TPP space. Putting together, these findings indicate that high-performing ACER agents can effectively identify dose violations from DVH inputs and employ a global tuning strategy to achieve high-quality treatment planning.

Significance: This study demonstrates that the AI agent learns effective TPP-tuning strategies, exhibiting behaviors similar to those of experienced human planners. Improved interpretability of the agent’s decision-making process may enhance clinician trust and inspire new strategies for automatic treatment planning.
\end{abstract}

\section{Introduction}
Automation in radiotherapy treatment planning represents a critical component of a fully automated workflow in radiation clinics. However, several challenges exist in this field, with one of the most critical being effective and efficient automatic inverse treatment planning. 

In inverse treatment planning, given clinicians' desired clinical outcomes such as tumor coverage and organ-at-risk (OAR) sparing, the treatment planning system (TPS) optimizes machine parameters, such as beam fluence maps, to meet these objectives. This process relies heavily on the proper configuration of treatment planning parameters (TPPs), which determine how different clinical goals and constraints are weighted during optimization. Traditionally, human planners adjust these TPPs through an iterative trial-and-error process, which is both time-consuming and labor-intensive. This is particularly a challenge in time-sensitive applications, such as online adaptive radiotherapy\citep{lim2017online,li2013automatic}. Over the past two decades, although there have been various efforts to automate this tuning process using either conventional algorithms or artificial intelligence (AI)-based strategies \citep{hussein2018automation, meyer2021automation, wang2019artificial, fu2021artificial}, widespread clinical adoption remains limited due to multiple challenges. \cite{meyer2021automation} reviewed multiple commercial solutions for automated planning and identified key drawbacks, including inflexibility in adapting to diverse dosimetric preferences, the requirement for large datasets to train AI models, and steep learning curves that hinder effective clinical usage. Furthermore, most existing solutions are site-specific, lacking generality across different anatomical sites or treatment techniques. AI-based solutions are typically associated with a black-box nature, leading to trust issues \citep{heising2023accelerating}.

To overcome these challenges, the key may lie in a deeper understanding of the inverse optimization problem itself. A better understanding of the TPP hyperspace and how individual TPP tuning contributes to trade-offs among competing dose objectives under different planning scenarios may help build generalized yet effective planning models applicable to various treatment sites and techniques. Yet, this is a complex task due to the multicriteria optimization nature of the problems\citep{zarepisheh2014multicriteria, craft2012improved}. While such complexity can be overwhelming for human planners, it is well within the capability of AI systems\citep{sahiner2019deep,shen2020introduction,shan2020synergizing}. As already demonstrated in the game of Go, deep reinforcement learning (DRL)-based system AlphaGo Zero developed a global perspective on sequences of moves, enabling it to produce high-quality moves across different board positions \citep{silver2017mastering}. Similarly, a DRL-based planning agent capable of learning globally optimal policies for TPP tuning could, in principle, be generalized to produce high-quality plans across diverse anatomical sites and treatment techniques.

In this regard, our groups and others have developed multiple DRL agents to operate TPSs for automatic treatment planning, achieving substantial initial progress. These DRL agents observe intermediate treatment plans generated by the TPS and automatically tune the TPPs to guide further optimization by the TPS \citep{shen2019intelligent, shen2020operating,  shen2021hierarchical, liu2022automatic, pu2022deep, sprouts2022development, wang2023integrated, yang2024automated, abrar2025actor, madondo2025patient}. However, the reward functions used to guide DRL agent convergence during training are often tied to specific dose objectives, and the underlying decision-making process remains difficult to interpret. These limitations create a gap between the learned policies and the generality and reliability needed for broad clinical application. 

To bridge this gap, we propose the use of explainable AI (EXAI) techniques \citep{saraswat2022explainable} to uncover the decision-making process of the DRL agent. In recent years, EXAI has seen growing application in cancer radiotherapy \citep{hou2024self, cui2023interpretable, teng2024literature, ladbury2022utilization, chatterjee2022torchesegeta}, particularly in interpreting AI models used for outcome prediction and image segmentation. For example, \cite{hosny2018deep} developed a CNN model to predict mortality risk in patients with non-small cell lung cancer (NSCLC) from CT images. Using the EXAI technique Grad-CAM \citep{selvaraju2017grad}, they visualized the tumor and surrounding dense tissue regions as key contributors to the model’s prediction. \cite{heising2023accelerating} further argued that EXAI can serve as a shared mental model between clinicians and AI systems, improving collaboration and facilitating integration into clinical workflows. However, the application of EXAI in sequential decision-making tasks such as DRL-based automatic treatment planning remains unexplored.

To investigate the factors underlying the DRL agent’s effectiveness, reliability, and efficiency in automatic planning, we conducted the following study. We selected our recently developed Actor-Critic with Experience Replay (ACER)–based DRL agent \citep{abrar2025actor} as the subject of analysis. \cite{abrar2025actor} demonstrated that training the ACER agent on a single patient case was sufficient to achieve both generalization and robustness, using prostate cancer intensity-modulated radiotherapy (IMRT) planning as the testbed. To gain further insights, we selected multiple checkpoint ACER agents from different stages of the training process and applied the integrated gradients (IG)–based EXAI method \citep{sundararajan2017axiomatic} to interpret how input states contributed to the agents’ TPP-tuning decisions. We then quantified each agent’s treatment planning performance in terms of both planning quality and efficiency. To connect the input attribution analysis with planning outcomes, we further analyzed the corresponding TPP tuning space. Based on these data, we performed a systematic analysis of how ACER agents make TPP-tuning decisions that lead to effective and efficient treatment planning. We found that a well-trained ACER agent can effectively identify dose violation regions from dose–volume histogram (DVH) inputs, distinguish the planning impact of different TPP-tuning actions, and promote those actions that globally reduce the magnitude of dose violations in an effective and efficient manner. To the best of our knowledge, this study represents the first application of EXAI in automatic treatment planning.

The remainder of this paper is organized as follows. We first provide a brief overview of the ACER agent for automatic treatment planning, the IG method, and the experimental setup used for evaluating and interpreting the agent. We then present our analysis results, followed by discussion and conclusion.

\section{Methods}
\subsection{The Actor-Critic with Experience Replay (ACER) Network for Automatic Treatment Planning}

\begin{figure}
    \centering
    \includegraphics[width=0.95\linewidth]{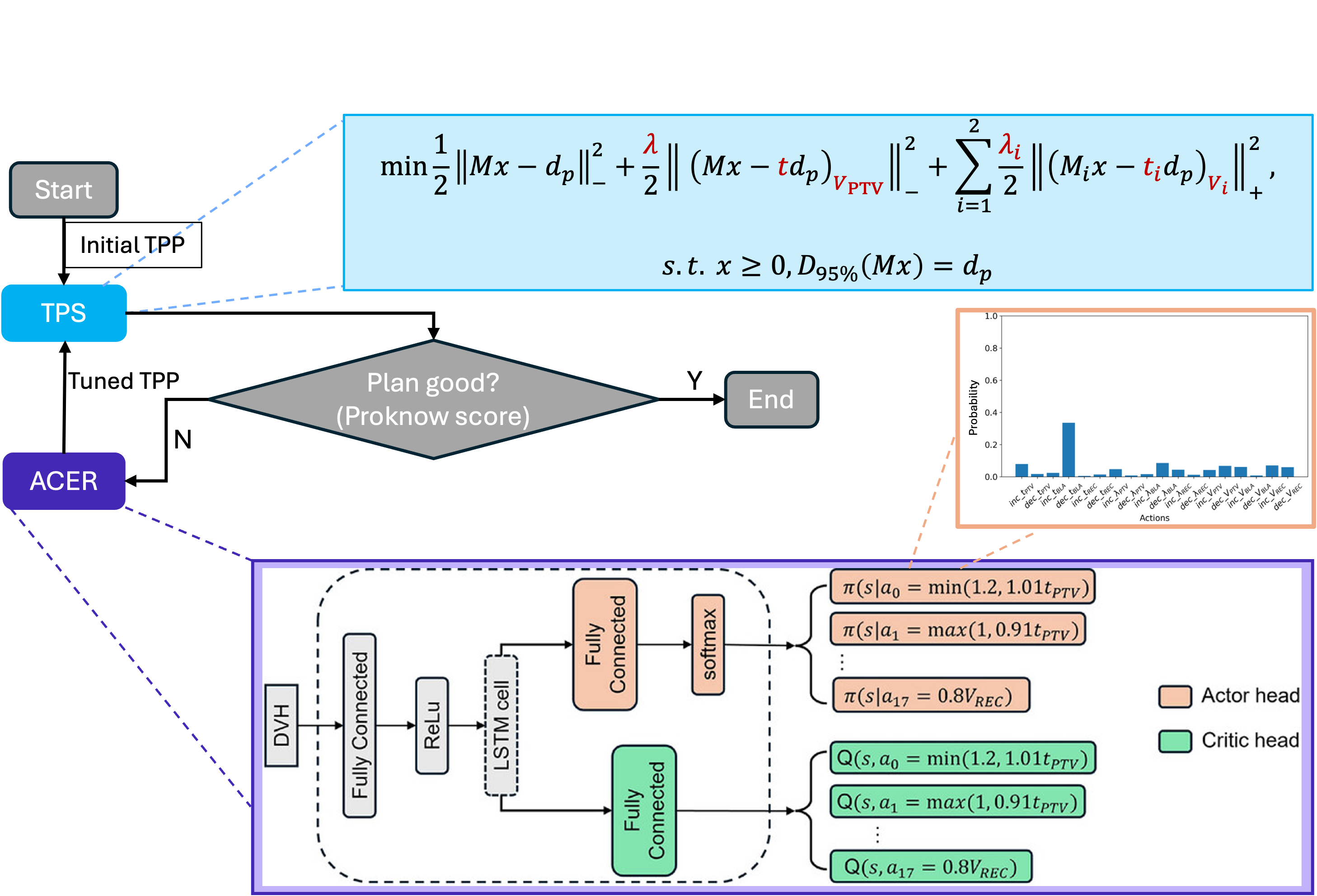}
    \caption{The illustration of the deep reinforcement learning (DRL) agent based automatic treatment planning for prostate cancer intensity modulated therapy.}
    \label{fig:planning system}
\end{figure}

Figure \ref{fig:planning system} illustrates the overall workflow of our ACER-based automatic treatment‐planning system for prostate-cancer IMRT. In the following, we briefly describe its main components. Interested readers may refer to our previous work \citep{abrar2025actor} for further details. 

In this workflow, we employed an in-house TPS whose fluence map objective function is shown in the light blue box of Figure \ref{fig:planning system}. There, $|\cdot|^2_-$ and $|\cdot|^2_+$ represent the under-dose and over-dose constraints, respectively. $M$ denotes the dose deposition matrix, $x$ is the beamlet vector, and $d_p$ is the prescription dose. The adjustable TPPs include the weighting factors $\lambda$ and $\lambda_i$, the upper dose constraints $t$ and $t_i$, and the volume constraints $V_{\mathrm{PTV}}$ and $V_i$. Here $i=1, 2$ corresponds to bladder and rectum, respectively. 

Starting from trivial initial TPPs, the TPS might fail to produce plans that meet clinical criteria (here, the ProKnow scoring system is employed). To address this, the ACER agent observed the dose-volume histogram (DVH) of the intermediate plan and generates TPP-tuning decisions. The tuned-TPPs were then fed back into the TPS for replanning. The process was iterated until a satisfying plan was obtained or the maximum number of iterations was reached.

The ACER agent itself employs a single-network, two-head architecture, with one head outputting Q-values and the other outputting policy distribution, as illustrated in the blue box. The specific policy space constitutes 18 TPP-tuning actions, which increases or decreases each TPP value by a fixed amount. A long short-term memory (LSTM) cell is embedded in the network's hidden layers to address training instabilities. This arises from the fact that the network receives only the DVH as user input, without access to the current TPPs or detailed evaluation metrics. As a result, the TPP tuning process forms a partially observed Markov decision process (POMDP), for which LSTM has proven effective in stabilizing its training process \citep{krishnamurthy2016partially, hausknecht2015deep, omi2023dynamic}.

\subsection{The Integrated Gradients Method}
Determining the contribution of DVH inputs to ACER’s policy distributions can be framed as an attribution problem. Consider a deep network represented by a function $F: \mathbb{R}^n \rightarrow [0, 1]$ with an input $x = (x_1, \ldots, x_n) \in \mathbb{R}^n$. The attribution problem seeks to quantify the contribution of each input feature $x_i$ to the prediction $F(x)$. This is defined relative to a baseline input $x'$, and an attribution method is supposed to assign an attribution vector $A_F(x, x') = (a_1, \ldots, a_n) \in \mathbb{R}^n$, where $a_i$ represents the contribution of feature $x_i$ to the output.

\cite{sundararajan2017axiomatic} proposed the IG method to obtain the attribution vector, with the IG value for the $i$th input $x_i$ to the $j$th output $F_j$ defined as
\begin{equation}\label{eq:IG}
\mathrm{IG}_{ij}(x, x', F_j) = (x_i - x'_i) \int_0^1 \frac{\partial F_j(x' + \alpha(x - x'))}{\partial x_i} \, \rmd \alpha.
\end{equation}
Here, $F$, $x$, and $x'$ follow the definitions provided in the attribution problem, and $\alpha \in [0, 1]$ represents the interpolation coefficient along the straight path from the baseline input to the actual input. 

The IG method uniquely satisfies two key axioms. The first is the sensitivity axiom, which states that input features causing different predictions should receive non-zero attributions. The second is the implementation invariance axiom, which requires that functionally equivalent networks produce identical attributions. Furthermore, the IG method satisfies the completeness property, which states that the sum of all attributions to one oneput, i.e., $\sum_{i=1}^{n} \mathrm{IG}_{ij}(x, x', F_j)$, equals the difference between $F_j(x)$ and $F_j(x')$  \citep{sundararajan2017axiomatic}, that is 
\begin{equation}\label{eq:IG sum}
\sum_{i=1}^{n} \mathrm{IG}_{ij}(x, x', F_j)=F_j(x)-F_j(x').
\end{equation}

\subsection{Experimental Setup for Attribution Analysis}
In our application, we used the IG method to examine how DVH inputs influenced the ACER agent’s TPP-tuning decisions. While this built a connection between DVH and TPP tuning, it did not fully explain the ACER agent’s overall treatment planning efficiency and effectiveness. To complete the picture, we first examined how DVH attributions aligned with TPP-tuning strategies. We then quantified the agent's treatment planning performance along with the associated TPP space and tuning behavior. By treating TPP tuning as a connecting bridge, we aimed to reveal how ACER effectively and efficiently performed automatic treatment planning in response to varying DVH inputs across different plans.

To support this analysis, we employed multiple checkpoint ACER agents saved at different stages of training. In our previous work \citep{abrar2025actor}, with one prostate cancer case, the ACER agent was trained for approximately 250,000 steps. The trained network showed full-score planning performance between 100,000 and 200,000 steps over two additional validation cases. A checkpoint agent at step 120,500 was used to report test performance in that study. 

In the present work, to interpret the policy space and tuning behavior, we selected six checkpoint agents from training steps 100,000, 120,500, 140,000, 160,000, 180,000, and 200,000. These agents are referred to as Agent\_10, Agent\_C, Agent\_14, Agent\_16, Agent\_18, and Agent\_20, respectively, throughout the study. Each agent was used to perform automatic treatment planning for 39 patient cases. These cases were drawn from the same dataset and followed the same TPP initialization protocol as Test Group 1 in Table 5 of our previous study \citep{abrar2025actor}, with the exception that we excluded cases that had already achieved the maximal ProKnow plan score at initialization.

\subsubsection{The Attribution from DVH and LSTM Inputs to ACER's TPP-tuning Decisions} \hfill\\
Before applying the IG method to our problem, it is important to revisit the LSTM cell to fully understand the structure of the ACER agent's inputs. The specific LSTM cell used in our implementation is defined as follows:
\begin{eqnarray}
\label{eq:LSTM}
f_t &=& \sigma(W_{if} x_t + U_{hf} h_{t-1} + b_f), \nonumber\\
i_t &=& \sigma(W_{ii} x_t + U_{hi} h_{t-1} + b_i), \nonumber\\
g_t &=& \tanh(W_{ig} x_t + U_{hg} h_{t-1} + b_g), \nonumber\\
o_t &=& \sigma(W_{io} x_t + U_{ho} h_{t-1} + b_o), \nonumber\\
c_t &=& f_t \odot c_{t-1} + i_t \odot g_t, \nonumber\\
h_t &=& o_t \odot \tanh(c_t).
\end{eqnarray}
In this equation, $f_t$, $i_t$, $c_t$, and $o_t$ represent the forget gate, input gate, cell state, and output gate, respectively. $h_t$ is the hidden state at time $t$. $W_{}$ and $U_{}$ denote the input and recurrent weight matrices, while $b$'s are the bias terms. The cell takes $x_t$, the input state at time $t$, and combines it with $h_{t-1}$ and $c_{t-1}$ from the previous time step, producing $h_t$, which is passed to the next layer and ultimately generates the policy distribution. From this structure, the ACER agent receives three inputs at each step: the user-provided $x_t$, and the internal hidden states $h_{t-1}$ and $c_{t-1}$ from the LSTM.

Given this input configuration, we applied the IG method to quantify each input's contribution to the agent’s policy decisions. In our application, we set the baseline values of all three inputs (DVH, $c$, and $h$) to zero. Defining the function $F$ in Equation~\ref{eq:IG} as the policy output for a specific TPP-tuning action, IG computed the attributions of the inputs to that action relative to the baseline. Per the completeness property of IG \citep{sundararajan2017axiomatic}, the sum of the attributions over each input quantifies how much that input source (DVH, $c$, and $h$) increases or decreases the action’s policy probability compared to the baseline. We applied this method to compute the attribution heatmaps of the three inputs for all TPP-tuning actions over all planning steps. 

To analyze the results, we conducted two complementary statistical studies:

First, we classified the attribution heatmaps into groups, each corresponding to a specific TPP-tuning action. For a given input type (DVH, $c$, or $h$), a heatmap was assigned to the group of the action with the highest total attribution. This grouping was performed separately for each input and each checkpoint agent to identify dominant attribution patterns across the policy space.

Second, we evaluated the relevance of DVH attributions by measuring their correlation with organ-wise immediate rewards. Organ-wise attribution was computed by summing the DVH input attributions for each organ. Corresponding rewards were calculated in two ways: (1) by summing the changes in Proknow scores for that organ before and after the action, and (2) by summing the changes in dose violations, defined as the volume excess of a DVH point over its corresponding Proknow score criterion. This produced a three-dimensional attribution vector and reward vector (one component per organ) for each planning step and action. We then computed the cosine similarity between the two vectors as:
\begin{equation}\label{eq:similarity}
\textrm{Similarity} = \frac{1}{N}\sum_{i=1}^{N}\sum_{j=1}^{18} p_{ij}\cos(\vec{A}_{ij},\vec{R}_{ij}).
\end{equation}
Here, $N$ is the total number of TPP tuning steps, and $p_{ij}$, $\vec{A}_{ij}$, and $\vec{R}_{ij}$ denote the probability, attribution vector, and reward vector at the $i$th planning step for the $j$th action. The normalization by $N$ ensures the similarity is comparable across agents, as they may take different numbers of steps to plan the same case.

\subsubsection{ACER's Automatic Planning Performance and Associated TPP Tuning Behavior} \hfill\\
To evaluate each agent’s planning effectiveness and efficiency, we conducted ACER-guided automatic treatment planning for all 39 patient cases. As shown in Figure~\ref{fig:planning system}, TPP-tuning decisions were made based on a probability distribution. To account for this stochastic nature of the decision-making process, each case was planned five times, resulting in 195 final treatment plans per agent. We then computed the means and standard deviations of the planning scores to assess effectiveness, and measured the number of planning steps per final plan to quantify efficiency.

In the above automatic treatment planning process for the 195 plans per agent, we recorded the policy distribution at each TPP-tuning step and the final TPP configuration at the termination step. Using these records, we performed two analyses with the aim to better understand how successful agents navigate the TPP space and how their policy distributions reflect their decision strategies.

First, to identify patterns in the TPP space, we selected the treatment plans that achieved the maximal Proknow score and plotted their final TPP configurations. To quantify the convergence efficiency of each configuration, we calculated the ``ideal'' number of planning steps, defined as the total number of monotonic, one-directional adjustments needed to reach the final TPP configuration from the initial settings.

Second, to characterize the agent’s decisiveness and strategy diversity, we calculated the entropy of the policy distribution per treatment plan as 
\begin{equation}\label{eq:entropy}
\textrm{Entropy} = -\frac{1}{N}\sum_{i=1}^{N}\sum_{j=1}^{18} p_{ij}\log_{10}(p_{ij}).
\end{equation}
Here, $p_{ij}$ is the probability of $j$th TPP tuning action at the $i$th TPP tuning step, and $N$ is the total number of TPP tuning steps. The normalization by $N$ ensures the entropy is comparable across agents. Higher entropy indicates a more uniform policy distribution across actions, whereas lower entropy reflects a more concentrated distribution toward specific actions.

\section{Results}
\subsection{The Attribution from DVH and LSTM inputs to ACER's TPP-Tuning Decisions}\label{sub:DVH attribution}
\begin{figure}
    \centering
    \includegraphics[width=1\linewidth]{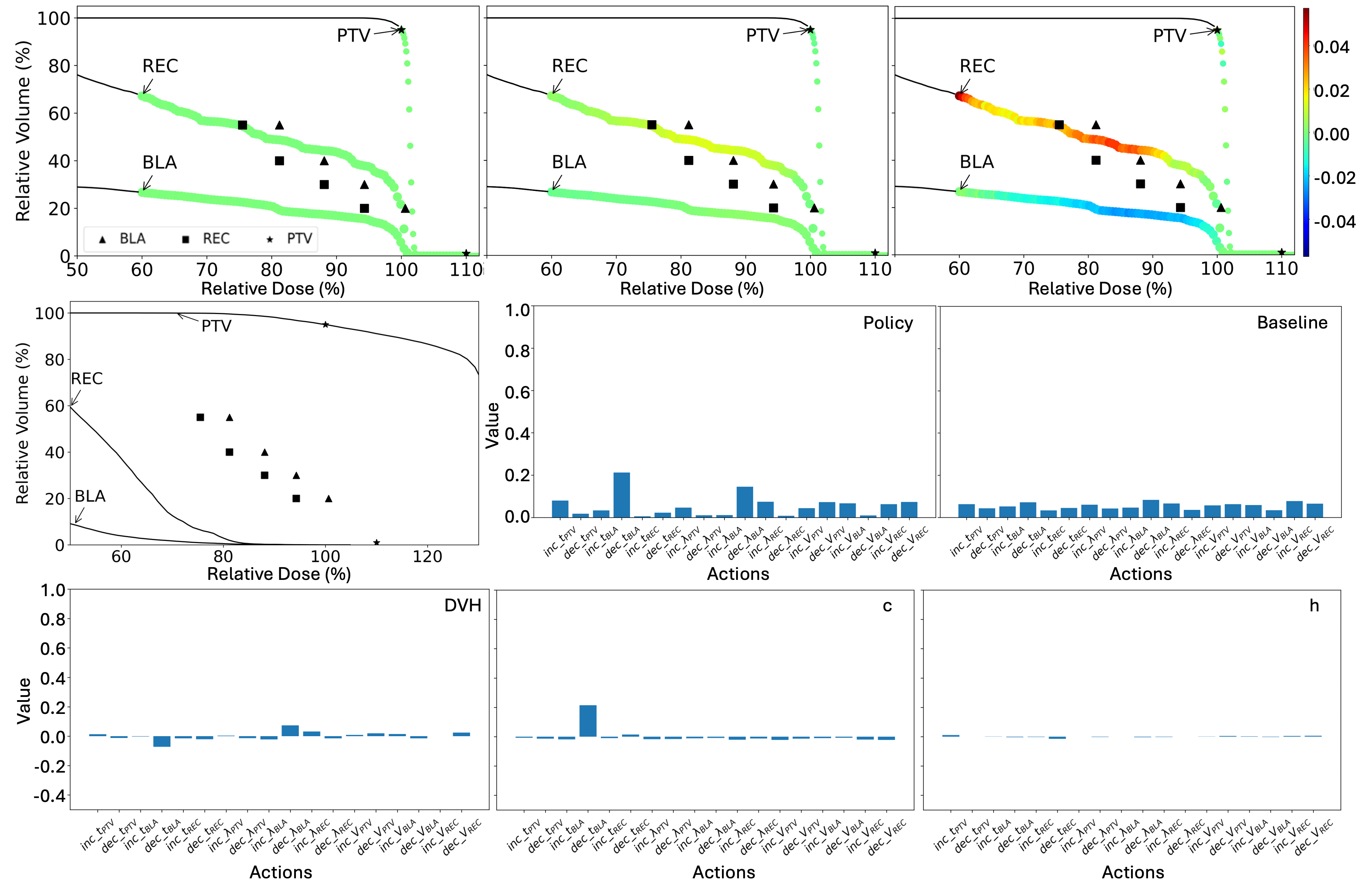}
    \caption{First row: The DVH IG heatmaps for the leading action in a representative treatment plan from three ACER agents. Second and third rows: the overall policy distribution across the baseline, DVH, and memory $c$ and $h$ components for a representative plan and agent.}
    \label{fig:DVH IGs}
\end{figure}

The DVH IG heatmaps contributing to a representative leading action in one TPP-tuning step for three ACER agents are shown in the first row of Figure \ref{fig:DVH IGs}. In this example, the plan fails to adequately spare the rectum, as indicated by the rectum curve exceeding the corresponding Proknow evaluation criteria (square markers). The three agents are taken from training step 0, a checkpoint before Agent\_10, and a checkpoint after. As training progresses, the rectum IG heatmap shifts from a neutral contribution (green) to leading action to a positive contribution (red). By the completeness property of IG, a larger total IG value corresponds to the leading action having a higher probability in the policy space. Thus, this change indicates that the agents gradually learn to detect dose violations and use this information to promote TPP-tuning actions.

The remaining two rows of Figure \ref{fig:DVH IGs} present the overall policy distribution across the baseline, DVH, and LSTM components ($c$ and $h$) for a representative agent and TPP-tuning step. The action probabilities derived from the baseline input are nearly uniform, indicating that zero inputs from both DVH and LSTM produce a flat policy distribution. In contrast, the DVH and LSTM components generate distinct and non-uniform policy distributions, highlighting the agent's sensitivity to non-zero DVH and LSTM inputs in its decision-making process.

\begin{figure}
    \centering
    \includegraphics[width=1\linewidth]{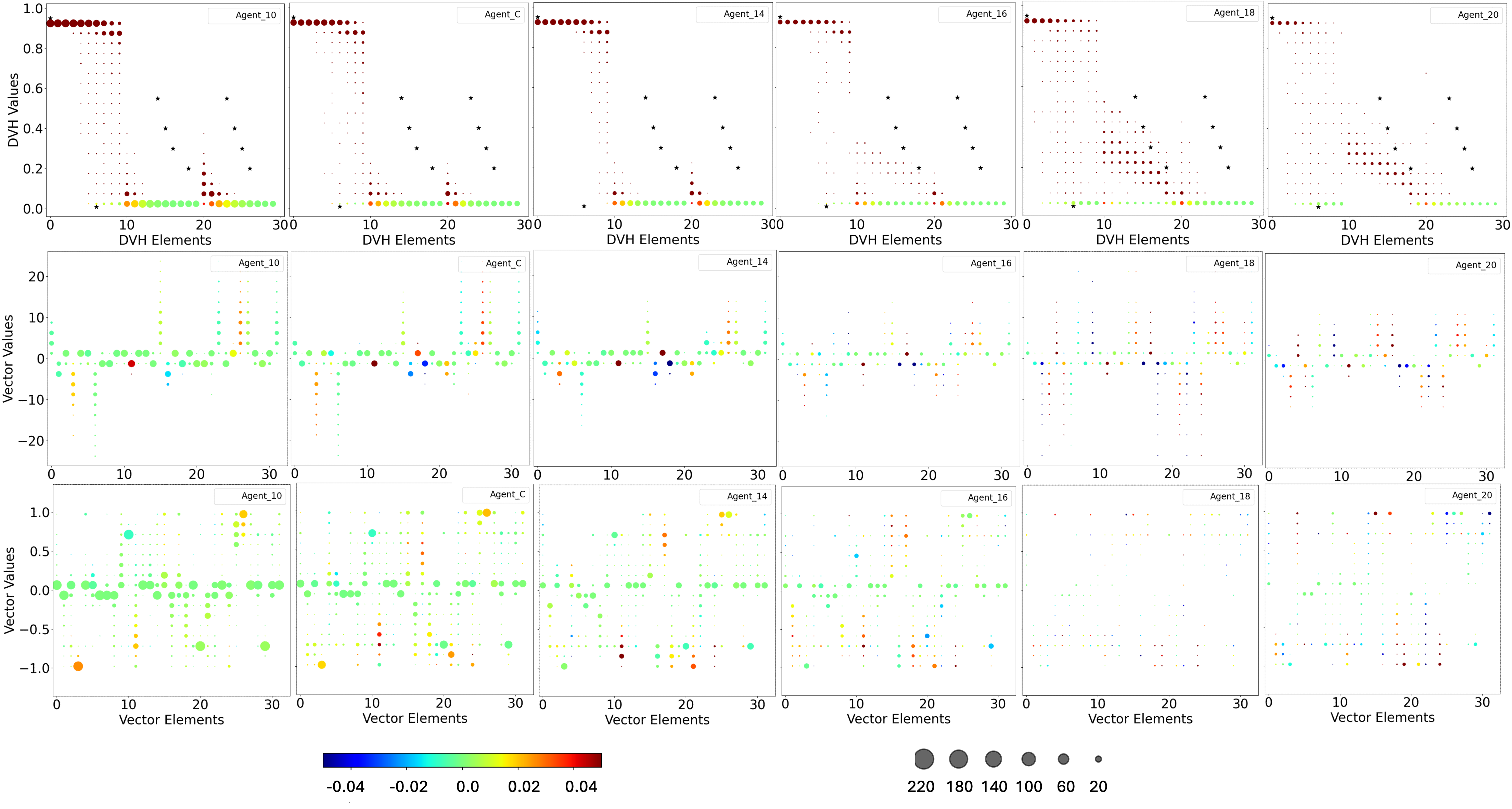}
    \caption{First row: Attribution statistics from DVH inputs to the leading action “decreasing $\lambda_{REC}$” across all six agents. Data were collected from all planning steps of test cases where this action received the highest total IG attribution from DVH inputs. The 300 original DVH input values were grouped into 30 segments of 10 consecutive points each: indices 0–9 represent the PTV, 10–19 the bladder, and 20–29 the rectum. The ProKnow score criteria used to evaluate the plan quality are marked with black stars. Second row: Corresponding attribution statistics from the LSTM memory cell state ($c$), consisting of 32 indices. Third row: Attribution statistics from the LSTM hidden state ($h$), also with 32 indices.}
    \label{fig:DVH c and h attributions}
\end{figure}

Figure \ref{fig:DVH c and h attributions} shows the IG heatmap statistics from DVH and LSTM inputs to a representative leading action, ''decreasing $\lambda_{REC}$''. IG values for each ACER agent were collected across 78 treatment plans, corresponding to two planning runs for each of 39 patient cases. ''decreasing $\lambda_{REC}$'' were identified as leading action from 398, 252, 212, 128, 160, and 91 planning steps for DVH; 316, 234, 206, 76, 75, and 69 steps for cell state $c$; and 634, 452, 299, 187, 19, and 80 steps for hidden state $h$, respectively. For DVH inputs, indices 0–10, 10–20, and 20–30 correspond to the PTV, bladder, and rectum curves, respectively. The ProKnow evaluation thresholds are marked with black stars. 

From the plot, PTV points typically lie above their evaluation criteria, indicating overdose, while OAR points are generally below. This implies that the agents tend to promote ''decreasing $\lambda_{REC}$'' action as leading action in situations with overdosed PTVs. Yet, moving from early to later agents, the level of DVH ``hot tail'' decreases, while the bladder shows more points close to violating the plan evaluation criteria. Correspondingly, the positive IG values shift from being primarily PTV-focused to reflecting contributions from both the PTV and bladder. Since decreasing $\lambda_{REC}$ reduces the rectum's planning weight and effectively increases the relative importance of the PTV and bladder, later agents appear to use this action to optimize dose distribution to both structures.
\begin{figure}
    \centering
    \includegraphics[width=1\linewidth]{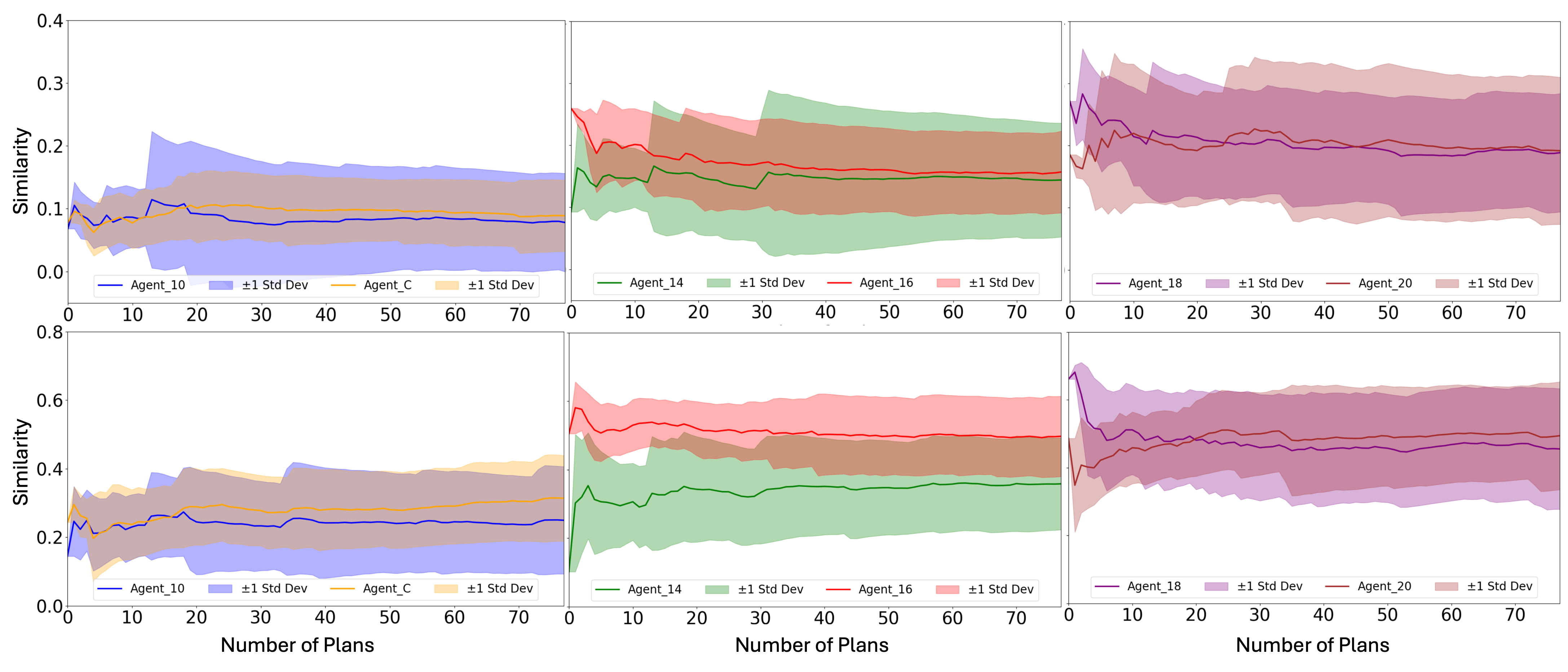}
    \caption{Similarity between organ-wise attribution and organ-wise immediate reward across the six ACER agents. The top row shows results using reward calculation Method 1 (score difference), while the bottom row uses Method 2 (dose violation difference).}
    \label{fig:similarity}
\end{figure}

The LSTM attributions from the cell state $c$ and hidden state $h$ reveal consistent patterns. First, although $c$ and $h$ have quite different value ranges for each agent, the vector components with positive contributions are largely similar. Second, in early agents, most $c$ values are near zero with occasional large spikes, whereas in later agents, they are more evenly distributed away from zero with fewer outliers. In parallel, the number of vector components with positive contributions increases. While the exact physical interpretation of the $c$ and $h$ components remains unclear, the observed trends align with those seen in the DVH inputs. Early spikes in $c$ may correspond to isolated extreme overdose events, such as in the PTV hot tail, whereas the broader non-zero $c$ values in later agents may reflect high, though not extreme, dose distributions in both the PTV and bladder. The greater number of $c$ and $h$ components with positive contributions in later agents also aligns with the increased positive DVH inputs from both structures.

Figure \ref{fig:similarity} gives the similarity between organ-wise attribution and organ-wise immediate reward for the six ACER agents. In the top row (Reward 1: score difference), Agents 18 and 20 exhibit the highest similarity (around 0.2), followed by Agents 14 and 16 (around 0.15), while Agents 10 and C remain below 0.1. In the bottom row (Reward 2: dose violation difference), all agents demonstrate much higher similarity values. Agents 16, 18, and 20 show the strongest alignment (around 0.5), with a noticeable gap from the remaining agents. The consistently higher and more distinguishable similarity under Reward 2 suggests that it better reflects the agents’ decision-making process. In particular, once an agent identifies an organ dose violation, the action it promotes within that TPP-tuning step may not immediately improve the plan score, but it can reduce the magnitude of that dose violation. With a well-aligned dose-violation observation and an action to mitigate it, the agent can, in principle, effectively improve the plan quality to ultimately achieve the highest score.

\subsection{ACER's Automatic Planning Performance and Associated TPP Tuning Behavior}\label{sub: TPP tuning}
Table \ref{tab:score and speed performance} presents statistical summaries of the planning effectiveness and efficiency for each ACER agent. From the table, all agents improve the plan qualities from a Proknow score of $5.15\pm1.72$ before planning to nearly the maximum of 9 after planning. However, there is a large variation among agents in the number of plans reaching the maximum Proknow score of 9, ranging from 143 for Agent\_10 to 191 for Agent\_16 out of 195 planning cases. The number of planning steps to get these full-score plans also varies substantially: Agent\_10 requires about $22.4\pm5.5$ steps for planning, while later agents, such as Agent\_18 generate full-score plans in nearly half the steps ($12.19\pm3.57$). Particularly, Agent\_16 achieves the highest plan score with the second-fastest planning speed, both with the smallest standard deviations, demonstrating stable, high performance.

\begin{table}[ht]
\centering
\caption{Treatment planning performance comparison among the six ACER agents. The initial and final plan scores were computed over 195 planning cases. The number of treatment plans (out of the 195) that achieved the maximum Proknow score of 9 is reported as ``\# of 9's''. The required number of planning steps and the ``ideal'' number of planning steps were calculated from this subset of plans.}
\begin{tabular}{m{1.8 cm}m{2.2 cm}m{2.2 cm}m{1.4 cm}m{2.5 cm}m{2.9 cm}}
\hline
Agent & Initial score & Final score &\# of 9's & \# of steps& Ideal \# of steps\\
\hline
Agent\_10 & \multirow[c]{6}{*}{$5.75 \pm 1.72$} & $8.75 \pm 0.44$ & 143&$22.35 \pm 5.46$& $14.97 \pm 3.10$\\
Agent\_C  & & $8.90 \pm 0.32$ & 176&$17.96 \pm 4.21$& $12.96 \pm 2.06$\\
Agent\_14 & & $8.98 \pm 0.16$ & 190&$16.01 \pm 2.89$& $13.03 \pm 1.74$\\
Agent\_16 & & $8.99 \pm 0.08$ & 191&$13.73 \pm 2.48$& $11.10 \pm 1.60$\\
Agent\_18 & & $8.93 \pm 0.30$ & 177&$12.19 \pm 3.57$& $9.38 \pm 2.07$\\
Agent\_20 & & $8.98 \pm 0.16$ &185 &$13.99 \pm 5.13$& $9.14 \pm 1.82$\\
\hline
\end{tabular}
\label{tab:score and speed performance}
\end{table}

\begin{figure}
    \centering
    \includegraphics[width=1\linewidth]{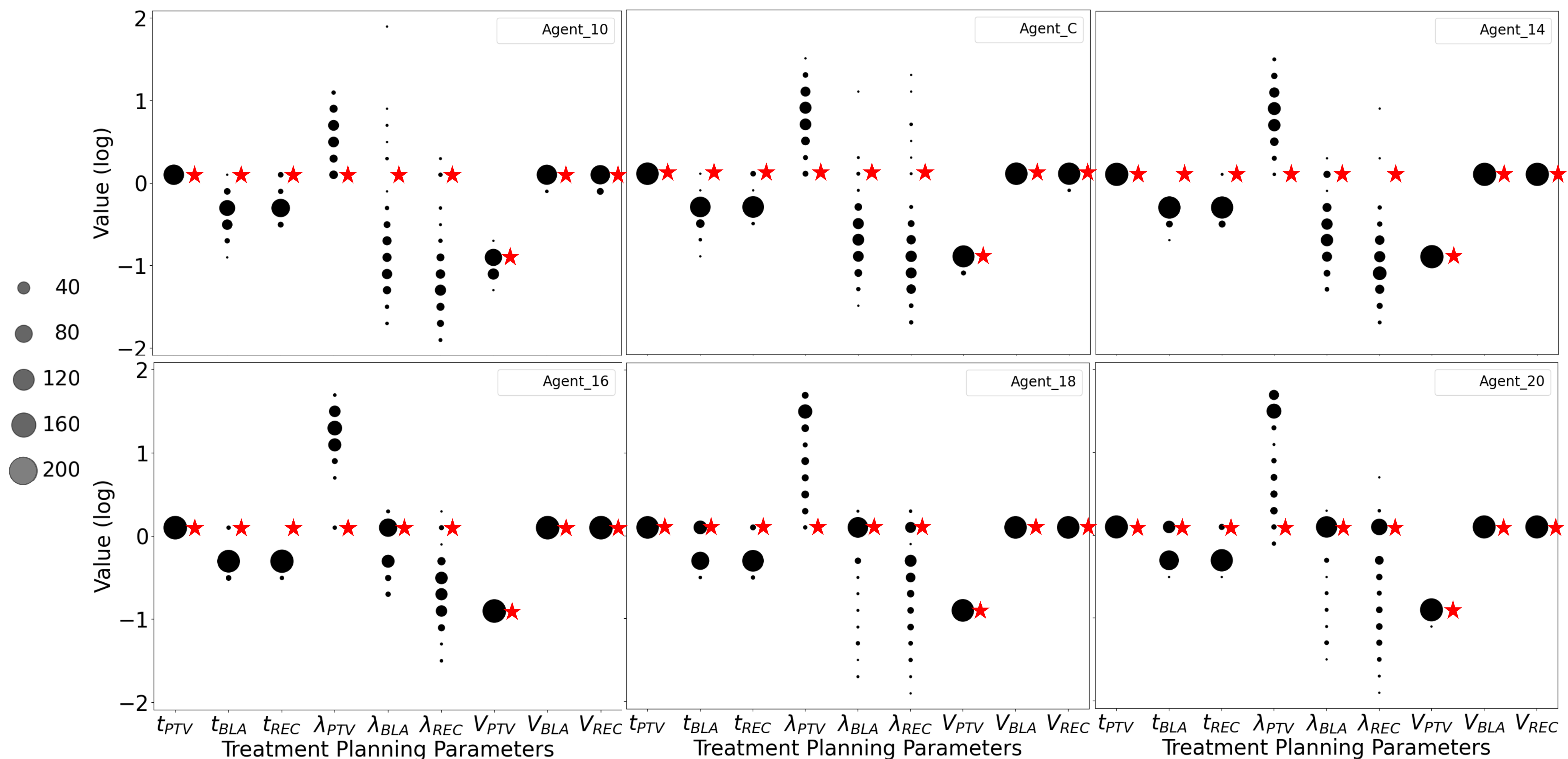}
    \caption{The value distributions in the treatment planning parameter (TPP) space corresponding to planning cases that achieved the maximum score of 9 for each of the six ACER agents. The red starred points indicate the initial TPP values prior to treatment planning, while the dotted points represent the final TPP values. The size of each dot is proportional to the number of planning cases that converged to that specific TPP value.}
    \label{fig:TPP space}
\end{figure}

Figure \ref{fig:TPP space} shows the final TPP space for plans that achieve a final planning score of 9. The six agents exhibit similar distribution patterns over the final TPP values. For example, the main tuning focus is on the weighting parameters $\lambda$'s, indicating shared tuning priorities. However, differences still exist. Comparing to other agents, Agent\_10’s dose thresholds $t_{REC}$ and $t_{BLA}$ converge to more distinct values. Both Agent\_10 and Agent\_C display bidirectional convergence in $\lambda_{REC}$ and $\lambda_{BLA}$, while the rest tend to converge in only one direction. Later agents converge to fewer distinct TPP points (e.g., the fewest dots in Agent\_16) and preserve more of their initial TPP settings, as shown by the larger black dots near the red-star markers (e.g., for $\lambda_{REC}$ in Agent\_16, Agent\_18 and Agent\_20). 

The ``ideal'' number of planning steps is then listed in the last column of Table \ref{tab:score and speed performance}, which decreases from around 15 in Agent\_10 to about 9 in Agent\_20. This trend is consistent with the final TPP space distribution patterns that later agents retain more initial TPP values than early agents. However, the actual planning steps (second-to-last column) remain higher than these values, reflecting back-and-forth TPP tunings during the planning process. Nonetheless, later agents exhibit a smaller discrepancy between the two measures, with Agent\_16 reaching a minimum.

\begin{figure}
    \centering
    \includegraphics[width=1\linewidth]{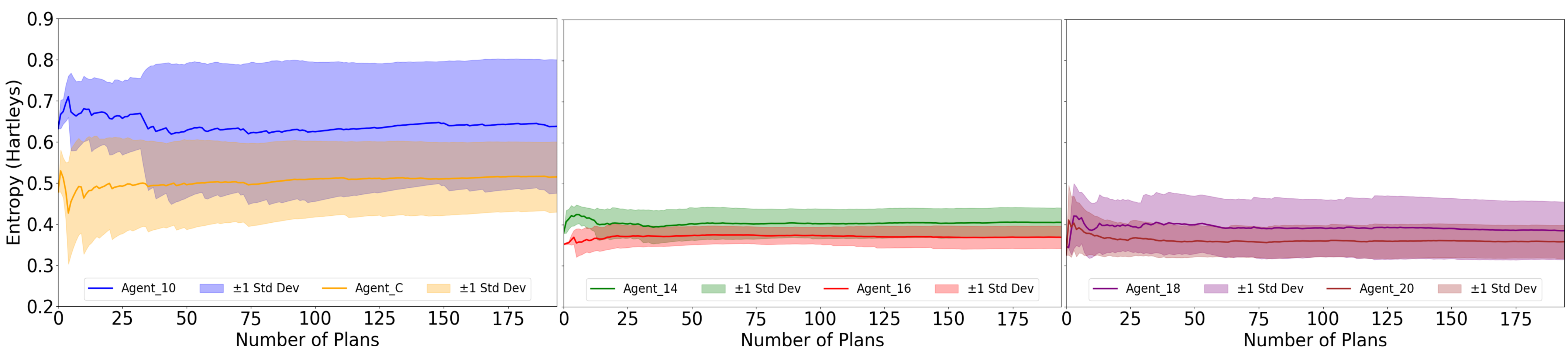}
    \caption{The entropy of the policy space plotted against the number of treatment plans for the six ACER agents.}
    \label{fig:entropy}
\end{figure}

The entropy of the policy space is illustrated in Figure \ref{fig:entropy}. Later agents display lower entropy with smaller standard deviation compared to earlier agents (e.g., Agent\_10, Agent\_C). Since lower entropy corresponds to more non-uniform distributions, this suggests that later agents have more dominant leading actions in the policy space.

\subsection{Overall Interpretation of ACER-based Automatic Planning} \label{sub:overall}

In this subsection, we connect the results from subsection \ref{sub:DVH attribution} with those from subsection \ref{sub: TPP tuning} to provide an overall interpretation of the decision-making process in ACER-based automatic treatment planning.

From Table \ref{tab:score and speed performance}, later agents such as Agent\_16 demonstrate superior performance in both planning quality and efficiency compared with earlier agents such as Agent\_10 and Agent\_C. The following paragraphs present our interpretation of this behavior.

From Figures \ref{fig:DVH IGs} and \ref{fig:DVH c and h attributions}, later agents show better recognition of actual and potential dose violations and display more positive IG distributions toward leading actions, thereby increasing the probability of these actions in the policy space. This is consistent with the entropy analysis in Figure \ref{fig:entropy}, which shows that later agents have more dominating leading actions. From Figure \ref{fig:similarity}, these promoted actions can effectively reduce the magnitude of dose violations. Combined with the smaller discrepancy between the “ideal” and actual numbers of planning steps for these agents as shown in Table \ref{tab:score and speed performance}, this indicates that the promoted actions have a global effect, systematically improving plan quality without back-and-forth adjustments of a single TPP parameter. After several TPP-tuning steps, this naturally produces high-quality treatment plans. The uncertainty levels in the similarity and entropy analyses also align well with those in Table \ref{tab:score and speed performance}, further supporting this reasoning.

Meanwhile, although later agents all show high effectiveness in achieving high final planning scores, their planning speeds and the number of plans that reach the maximum Proknow score vary. The variation can be interpreted as follows. The final TPP space distribution (Figure \ref{fig:TPP space}) shows that the promoted actions are primarily ``increasing $\lambda_{PTV}$'', ``reducing $\lambda_{REC}$'', and ``reducing $\lambda_{BLA}$''. Because only the relative values of the weighting parameters $\lambda$'s matter in the inverse optimization problem(Figure \ref{fig:planning system}), increasing $\lambda_{PTV}$ is effectively equivalent to reducing both $\lambda_{REC}$ and $\lambda_{BLA}$. Agents 14 to 20 all mainly focus on tuning these $\lambda$ values, thus achieving similarly high-quality treatment planning. However, as training progresses from Agent\_14 to Agent\_20, the focus shifts toward solely increasing $\lambda_{PTV}$, which shortens tuning steps and improves efficiency. In some situations, though, individually tuning these $\lambda$'s are important. This strategy is used by Agent\_16, which achieves the highest planning effectiveness and the second-fastest planning speed. By contrast, later agents such as Agent\_18 and Agent\_20 rely too heavily on solely increasing $\lambda_{PTV}$. This approach increases their planning speed but reduces flexibility, resulting in fewer maximum-score plans compared to Agent\_16.  

\section{Discussion}
In this study, we examined the attributions underlying the ACER agent's intelligent performance in automatic treatment planning. We found that an ACER agent with effective and efficient planning behavior can identify dose violations from DVH inputs and promote proper TPP-tuning actions to reduce the violation magnitude. After several tuning steps, these actions result in the generation of high-quality treatment plans. Although the treatment planning scenario used in this study was relatively simple, involving only one PTV and two OARs, the conclusions drawn from it can still offer meaningful insights into practical treatment planning by clinical TPSs. 

From a naive perspective, it can be difficult to distinguish the effects of tuning different TPPs on improving plan quality. Yet, the ACER agent learns to differentiate their relative importance through training. As shown in Figure \ref{fig:TPP space}, the agent discovers that by primarily adjusting the $\lambda$ parameters, plan quality can be improved both effectively and efficiently. In Figure \ref{fig:DVH c and h attributions}, when dose violations appear in both the PTV and one OAR, the agent chooses to reduce the $\lambda$ for the other OAR, which paradoxically helps mitigate both violations. This behavior leads to a global improvement in plan quality, achieving the highest ProKnow score without exhibiting back-and-forth tuning. These actions can thus be regarded as high-quality TPP-tuning strategies. Interestingly, in clinical practice, experienced human planners also tend to tune only a few key parameters with minimal tuning steps. The AI agent appears to acquire similar experience during training, much like a human. 

Inspired by these findings, we plan to extend the ACER-based auto-planning framework to more complex treatment planning scenarios, allowing it to learn more sophisticated TPP-tuning strategies when facing trade-offs among a larger number of dose objectives. This is feasible, as the ACER framework is highly scalable. We will then apply EXAI techniques to interpret the newly trained ACER agent, obtaining deeper insights into the TPP tuning space and the agent’s decision-making process. Through this training and interpretation process, we expect to build a DRL agent that develops transferable TPP-tuning insights, thus achieving generality and reliability in real-world clinical applications. In addition, the acquired knowledge could also assist the building of other AI models for effective planning.

Meanwhile, as demonstrated in the game of Go, the DRL-based AlphaGo Zero agent not only defeated top human players but also played an educational role by helping human players develop novel moves in challenging games \citep{shin2023superhuman}. Similarly, by quantitatively unraveling the black box of TPP-tuning effectiveness in complex inverse treatment planning, this knowledge can be used for educational purposes to train human planners. This understanding could also support the development of a more interactive human–AI environment, where attribution analysis illustrates the AI’s decision-making process and human planners can intervene if an unreasonable decision is detected.

\begin{figure}
    \centering
    \includegraphics[width=1\linewidth]{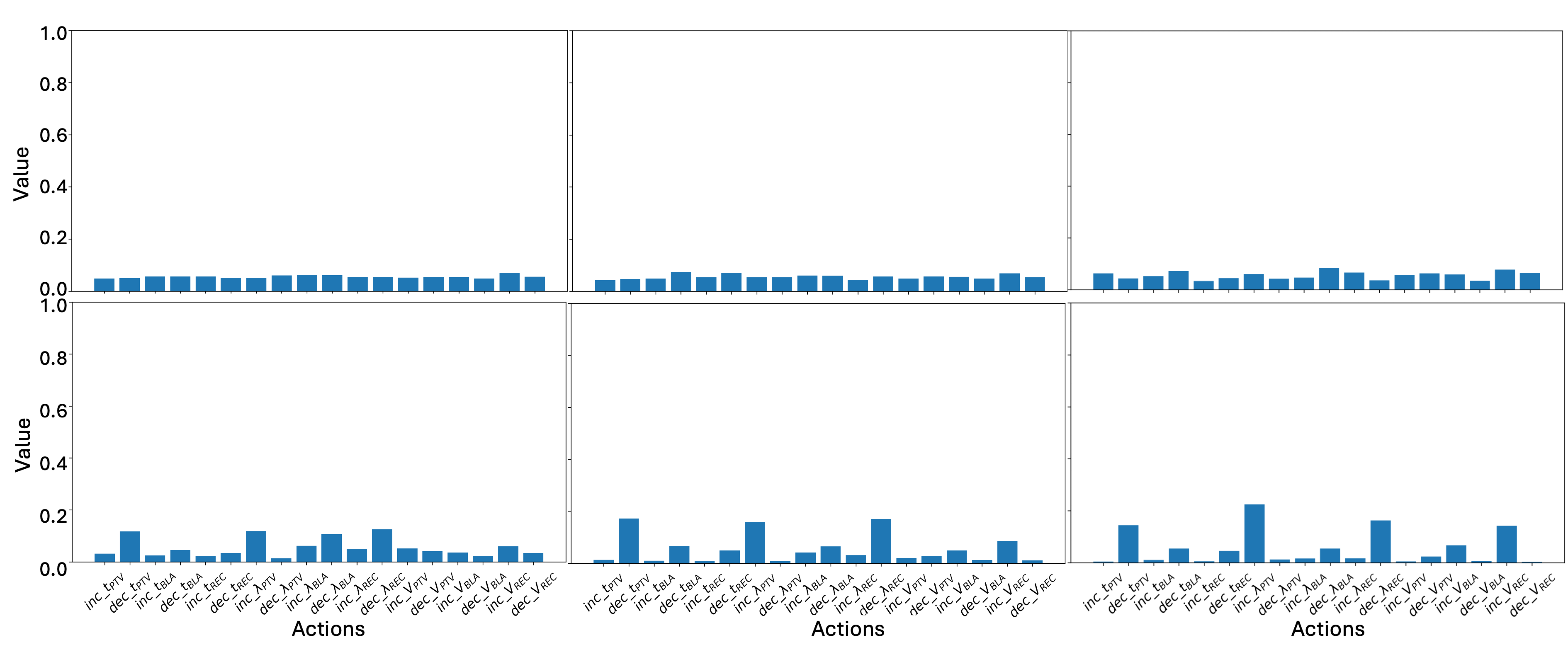}
    \caption{The baselines for the same DVH input obtained for ACER agents at training steps 0, 40,000, 80,000, 120,000, 160,000, and 200,000.}
    \label{fig:baseline}
\end{figure}

Nonetheless, except for the discussion of clinical impact, we would also like to address some technical considerations in the current EXAI study and outline a plan for future improvements. In this study, we used zero inputs as the baseline, which offered a convenient and uniform background at the start of training. However, as training progresses, later agents encountered different sets of treatment plans compared to earlier agents, resulting in a biased background when using zero inputs, as illustrated in Figure \ref{fig:baseline}. This violates the assumption of baseline neutrality and may influence the quantitative values in the DVH attribution analysis. Despite this, we argue that it does not significantly affect the overall attribution trends, since the background-induced actions are not dominant, as shown in Figure \ref{fig:TPP space}. To achieve a more uniform and neutral background across agents, dynamic baseline inputs may be needed, following methods such as those described in \cite{morasso2025guidelines}. An unbiased baseline would not only enable more precise input attribution but may also reveal inherent properties of the trained agents, contributing to a deeper understanding of their behavior. We plan to investigate this direction in future work.

\section{Conclusion}
In this study, we performed attribution analysis to understand the decision-making process in ACER-based automatic treatment planning. We found that well-trained ACER agents can effectively identify dose violation regions on DVH inputs and promote appropriate TPP-tuning actions to address these violations. This leads to the discovery of efficient TPP-tuning paths to achieve full-score plans, thereby achieving high efficacy and efficiency in treatment planning.

\section{Acknowledgment}
This work was partially supported by the Rising Stars Fund from The University of Texas System and by the National Institutes of Health/National Cancer Institute (NIH/NCI) under grant numbers R01CA237269, R01CA254377, and R37CA214639.
\bibliography{./references1}
\end{document}